\documentclass[%
preprint,
superscriptaddress,
 amsmath,amssymb,
 aps,
]{revtex4-1}

\usepackage{graphicx}% Include figure files
\usepackage{dcolumn}% Align table columns on decimal point
\usepackage{bm}% bold math
\usepackage{color} % for writing
\usepackage{xspace}
\usepackage{multirow}
\usepackage{mathrsfs}

\begin{document}

%%%%%% ============new command============== %%%%%%
\newcommand{\hl}{\textcolor{red}}
\newcommand\sss{\scriptscriptstyle}
\newcommand{\mW}{m_{\sss W}}
\newcommand{\sW}{s_{\sss W}}
\newcommand{\cW}{c_{\sss W}}
\newcommand{\hc}{+\,\mathrm{h.c.}}
\newcommand{\ca}{\ensuremath{\bar c_{\sss \gamma}}\xspace}
\newcommand{\chw}{\ensuremath{\bar c_{\sss HW}}\xspace}
\newcommand{\chb}{\ensuremath{\bar c_{\sss HB}}\xspace}
\newcommand{\Hg}{\ensuremath{H\mathrm{+}\gamma}\xspace}
\newcommand{\mHg}{\ensuremath{m_{H\gamma}}\xspace}
\newcommand{\pT}{\ensuremath{p_{\text{T}}}\xspace}
\newcommand{\ifb}{\mbox{fb$^{-1}$}\xspace}
\newcommand{\GeV}{\ensuremath{\mathrm{GeV}}}
\newcommand{\TeV}{\ensuremath{\mathrm{TeV}}}
%%%%%% ============new command============== %%%%%%

\preprint{Submitted to Chinese Physics C}

\title{Constraining the anomalous Higgs boson coupling\\ in $H$+$\gamma$ production}
\thanks{Supported by
	       the National Natural Science Foundation of China (11875278),
           Beijing Municipal Science \& Technology Commission (Z181100004218003),
           the National Key Research and Development Program of China (2018YFA0404001),
           and the Hundred Talent Program of the Chinese Academy of Sciences (Y6291150K2). }%

\author{Liaoshan Shi}
\email[]{Liaoshan.Shi@cern.ch}
\affiliation{%
 School of Physics, Sun Yat-sen University, Guangzhou 510275, China\\
}%

\author{Zhijun Liang}
\author{Bo Liu}
\affiliation{
 Institute of High Energy Physics, Chinese Academy of Science, Beijing 100049, China\\
 }
\affiliation{
 Physical Science Laboratory, Huairou National Comprehensive Science Center, Beijing 101400, China\\
}%

\author{Zhenhui He}
\affiliation{%
 School of Physics, Sun Yat-sen University, Guangzhou 510275, China\\
}%


\begin{abstract}
  Higgs boson production in association with a photon ($H$+$\gamma$) offers a promising channel to test the Higgs boson to photon coupling at various energy scales. Its potential sensitivity to anomalous couplings of the Higgs boson has not been explored with the proton-proton collision data. 
  In this paper, we reinterpret the latest ATLAS $H$+$\gamma$ resonance search results within the Standard Model effective field theory (EFT) framework, using 36.1~fb$^{-1}$ of proton-proton collision data recorded with the ATLAS detector at $\sqrt{s}=13$ TeV. Constraints on the Wilson coefficients of dimension-six EFT operators related to the Higgs boson to photon coupling are provided for the first time in the $H$+$\gamma$ final state at the LHC.

\end{abstract}

\maketitle

%\tableofcontents

%%%%%%%%%%
\section{Introduction}
%%%%%%%%%%

After the discovery of the Higgs boson~\cite{HIGG-2012-27,CMS-HIG-12-028}, measurements of the Higgs boson couplings to the other fundamental particles became crucial tests of the nature of the Higgs boson.
In the Standard Model (SM), coupling of the Higgs boson to photon is forbidden at the tree level, 
and is induced by heavy particle loops in, e.g., $H \to \gamma \gamma$ and $H \to Z\gamma$ processes.
The Higgs-photon coupling has been extensively studied in the various Higgs boson decay channels including $H \to \gamma \gamma$ and $H \to ZZ^{*}/Z\gamma^{*}/\gamma^{*}\gamma^{*} \to 4\ell$ with the LHC data recorded by the ATLAS and CMS experiments~\cite{HIGG-2015-02,Aaboud:2018xdt,Aaboud:2017vzb,ATL-PHYS-PUB-2017-018,CMS-HIG-14-018,CMS-HIG-15-002,CMS-HIG-17-011,Englert:2015hrx,Ellis:2018gqa}.

Apart from the Higgs boson decay channels involving photons, Higgs boson production in association with a photon can also be used to measure the Higgs-photon coupling.
The \Hg production cross section is predicted to be very small in SM, but anomalous couplings introduced in models beyond SM (BSM) can have significant effects.
The \Hg process has been considered as a promising and clean channel at LEP~\cite{Djouadi:1996ws,PhysRevD.52.3919}, and was used by the DELPHI collaboration to search for anomalous couplings of the Higgs boson to vector bosons~\cite{Abreu:1999vt}.
At the LHC, potential sensitivity of the $pp\to\Hg$ process to anomalous Higgs-photon couplings has been discussed in Ref.~\cite{Khanpour:2017inb}. 
It is predicted that some Wilson coefficients of dimension-six operators related to Higgs-photon couplings can be probed down to $10^{-2}$ with 300 \ifb of $pp$ collision data at 14 TeV.
There is no particular analysis measuring anomalous Higgs-photon couplings via this channel using the LHC data.
The ATLAS and CMS collaborations have reported the results of heavy \Hg resonance searches in 13 TeV $pp$ collision data~\cite{Aaboud:2018fgi,CMS-PAS-EXO-17-019}. Apart from the resonance models, their results are also sensitive to non-resonant \Hg production and to the anomalous coupling between the Higgs boson and photon.
However, these results have not been interpreted as limits on the anomalous Higgs-photon coupling.

In this paper, the latest \Hg resonance search results from the ATLAS collaboration~\cite{Aaboud:2018fgi} are reinterpreted within the SM effective field theory (EFT), and are presented as constraints on the Wilson coefficients of dimension-six EFT operators.
The study is based on a $pp$ collision dataset of 36.1~\ifb at $\sqrt{s}=13$ TeV.

This paper is organized as follows. Section~{\ref{sec:eft}} gives a short overview of the EFT framework and a brief description of the signal Monte Carlo generation for the reinterpretation.
Section~{\ref{sec:analysis}} describes the analysis strategy. 
Section~{\ref{sec:result}} presents the constraints on the Wilson coefficients of dimension-six EFT operators that are obtained in the $H+\gamma$ channel, and compares them with the existing results from other measurements.
Our conclusions are summarized in Section~{\ref{sec:conclusion}}.

%%%%%%%%%%
\section{The effective field theory}
\label{sec:eft}
%%%%%%%%%%
In the SM effective field theory approach, the effects of BSM interactions are parametrized using higher-dimension operators in addition to the SM Lagrangian.
Leading contributions at collider energies are expected to originate from dimension-six operators. A general effective Lagrangian with dimension-six operators ${\cal O}_i$ takes the form
\begin{equation}
  {\mathscr L_{\rm eff}} = {\mathscr L}_{\rm SM}  + \sum_i \bar c_i {\cal O}_i\ .
\end{equation}
with the Wilson coefficients $\bar c_i$ describing the strengths of the BSM interactions.

We focus on a set of dimension-six operators known as the strongly-interacting light Higgs (SILH) Lagrangian~\cite{Giudice:2007fh}. It is written as 
\begin{eqnarray}
  {\mathscr L}_{\rm SILH} =&&\
    \frac{\bar c_{\sss H}}{2 v^2} \partial^\mu\big[\Phi^\dag \Phi\big] \partial_\mu \big[ \Phi^\dagger \Phi \big]
  + \frac{\bar c_{\sss T}}{2 v^2} \big[ \Phi^\dag {\overleftrightarrow{D}}^\mu \Phi \big] \big[ \Phi^\dag {\overleftrightarrow{D}}_\mu \Phi \big] 
  - \frac{\bar c_{\sss 6} \lambda}{v^2} \big[\Phi^\dag \Phi \big]^3 \nonumber\\
  &&\  - \bigg[
     \frac{\bar c_{\sss u}}{v^2} y_u     \Phi^\dag \Phi\ \Phi^\dag\cdot{\bar Q}_L u_R
   + \frac{\bar c_{\sss d}}{v^2} y_d     \Phi^\dag \Phi\ \Phi {\bar Q}_L d_R
   + \frac{\bar c_{\sss l}}{v^2} y_\ell\ \Phi^\dag \Phi\ \Phi {\bar L}_L e_R
   + {\rm h.c.} \bigg] \nonumber\\
  &&\
  + \frac{i g\ \bar c_{\sss W}}{\mW^2} \big[ \Phi^\dag T_{2k} \overleftrightarrow{D}^\mu \Phi \big]  D^\nu  W_{\mu \nu}^k
  + \frac{i g'\ \bar c_{\sss B}}{2 \mW^2} \big[\Phi^\dag \overleftrightarrow{D}^\mu \Phi \big] \partial^\nu  B_{\mu \nu} \\
  &&\
  + \frac{2 i g\ \bar c_{\sss HW}}{\mW^2} \big[D^\mu \Phi^\dag T_{2k} D^\nu \Phi\big] W_{\mu \nu}^k
  + \frac{i g'\ \bar c_{\sss HB}}{\mW^2}  \big[D^\mu \Phi^\dag D^\nu \Phi\big] B_{\mu \nu} \nonumber\\
  &&\
  +\frac{g'^2\ \bar c_{\sss \gamma}}{\mW^2} \Phi^\dag \Phi B_{\mu\nu} B^{\mu\nu}
   +\frac{g_s^2\ \bar  c_{\sss g}}{\mW^2} \Phi^\dag \Phi G_{\mu\nu}^a G_a^{\mu\nu}\ , \nonumber
\end{eqnarray}
where $W_{\mu \nu}^k$, $B_{\mu \nu}$ and $G_{\mu \nu}^a$ are the gauge field strength tensors, and $\Phi$ is the Higgs doublet.
Among all Wilson coefficients in the SILH Lagrangian, \ca, \chw and \chb are related to the anomalous Higgs-photon coupling through a direct $HZ\gamma$ or $H\gamma\gamma$ vertex.
With the presence of these BSM vertices, additional tree level diagrams, in particular an $s$-channel diagram via a virtual photon or Z boson as the mediator, can contribute to the $pp\to\Hg$ process and lead to a large relative change in its production cross section.
Therefore, the \Hg process is a sensitive probe for exploring the anomalous Higgs-photon coupling~\cite{Khanpour:2017inb}.

A public implementation of the SILH Lagrangian is available in the general Higgs Effective Lagrangian (HEL)~\cite{Alloul:2013naa,Contino:2013kra}. The HEL model is implemented in \textsc{FeynRules}~\cite{Alloul:2013bka}, comprising 39 dimension-six operators and their corresponding Wilson coefficients. Its Universal \textsc{FeynRules} Output~\cite{Degrande:2011ua} has been interfaced to the \textsc{MadGraph5}\_aMC@NLO~\cite{Alwall:2014hca} event generator.
In this work, the HEL model is used with all the other Wilson coefficients fixed to 0 except \ca, \chw and \chb.
The $pp\to\Hg$ production cross section is computed for different values of \ca, \chw and \chb, using \textsc{MadGraph5}\_aMC@NLO v2.6.2 with NNPDF2.3~\cite{Ball:2012cx} parton distribution functions.
We then parametrize the signal cross section as a function of the Wilson coefficients that result from the computation.
Figure~\ref{fig:xsec_2d} presents a two-dimensional parametrization of the signal cross section
parametrized as a function of two of the three Wilson coefficients, with the third coefficient fixed to 0.
Monte Carlo event samples are also generated with the same configurations.

\begin{figure}[hbt]
\includegraphics[width=0.32\textwidth]{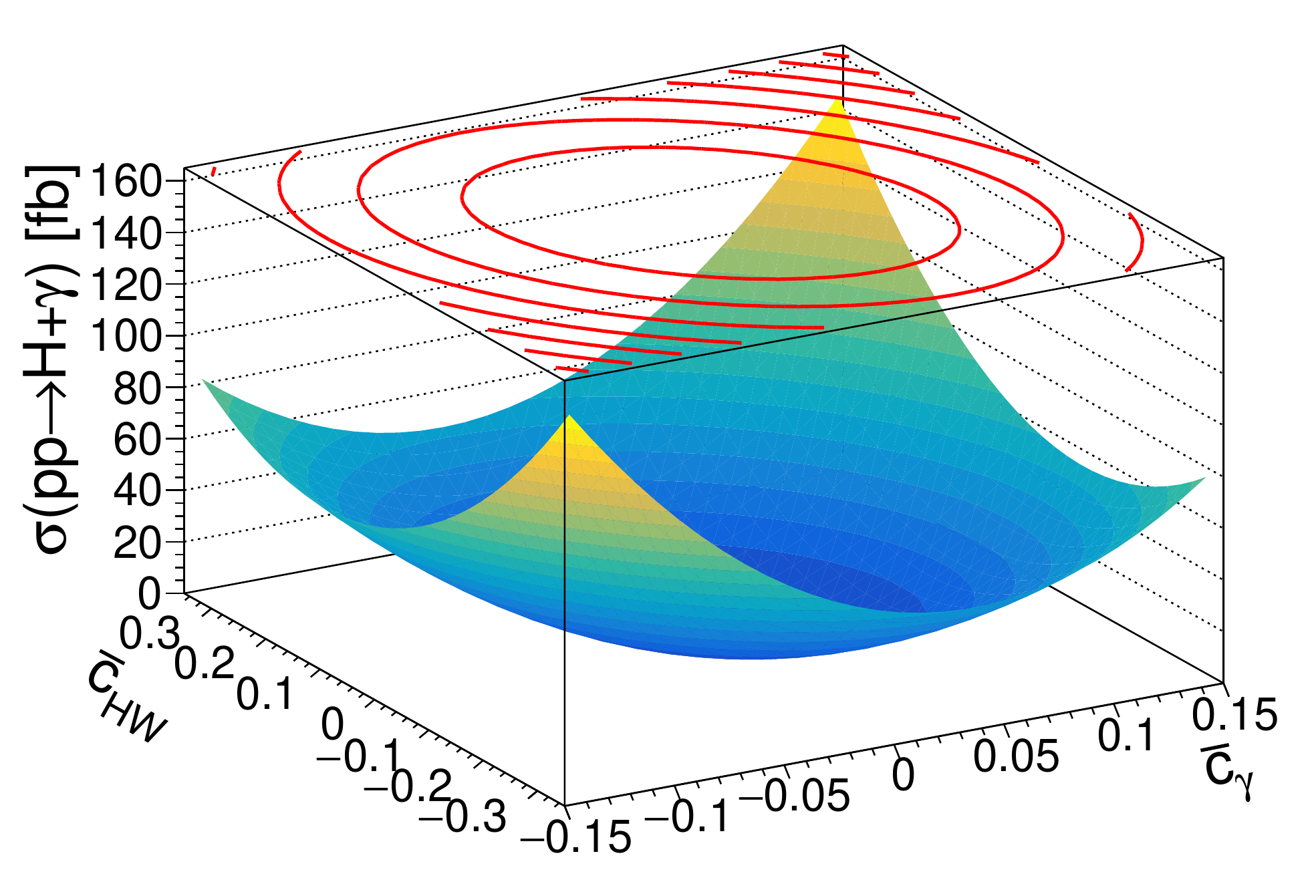}
\includegraphics[width=0.32\textwidth]{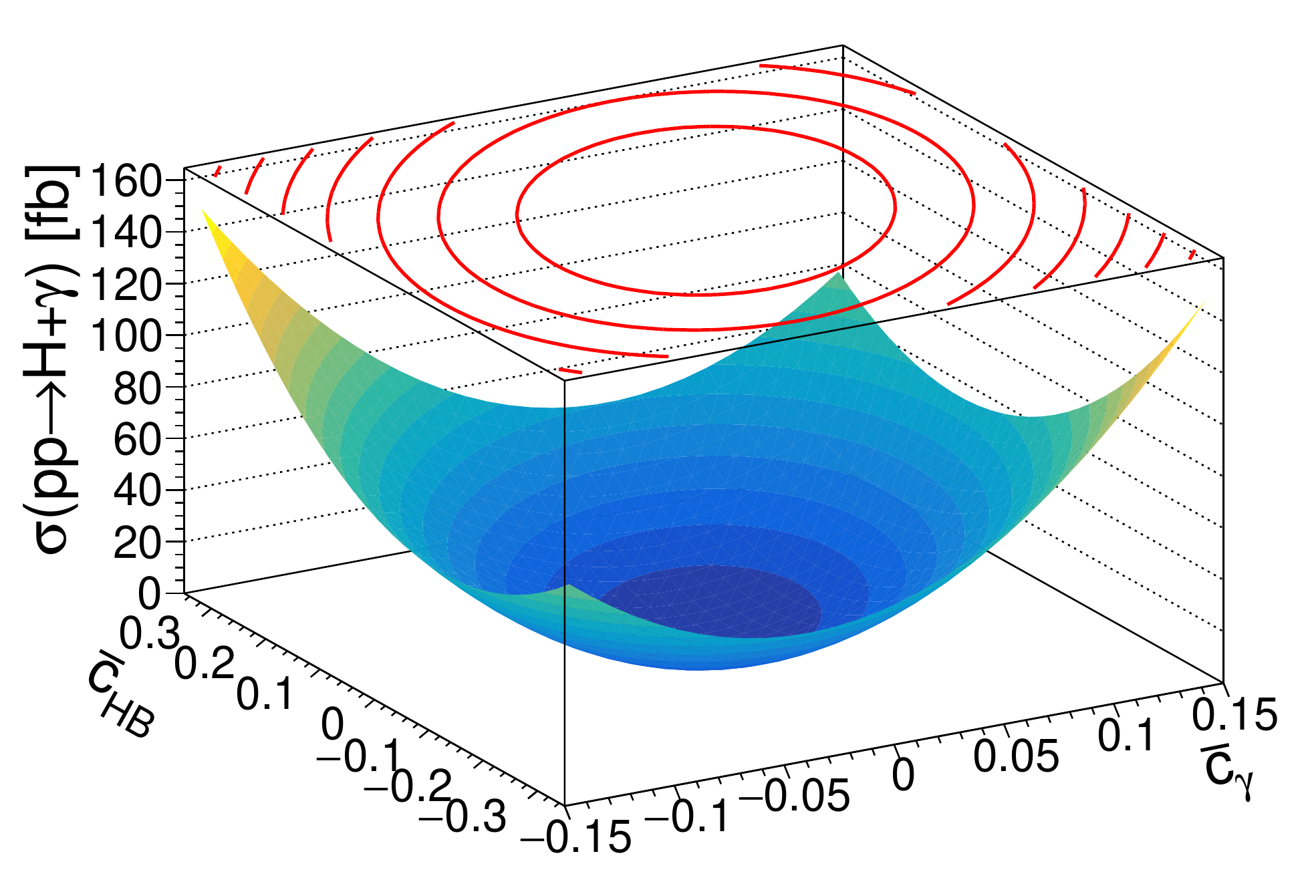}
\includegraphics[width=0.32\textwidth]{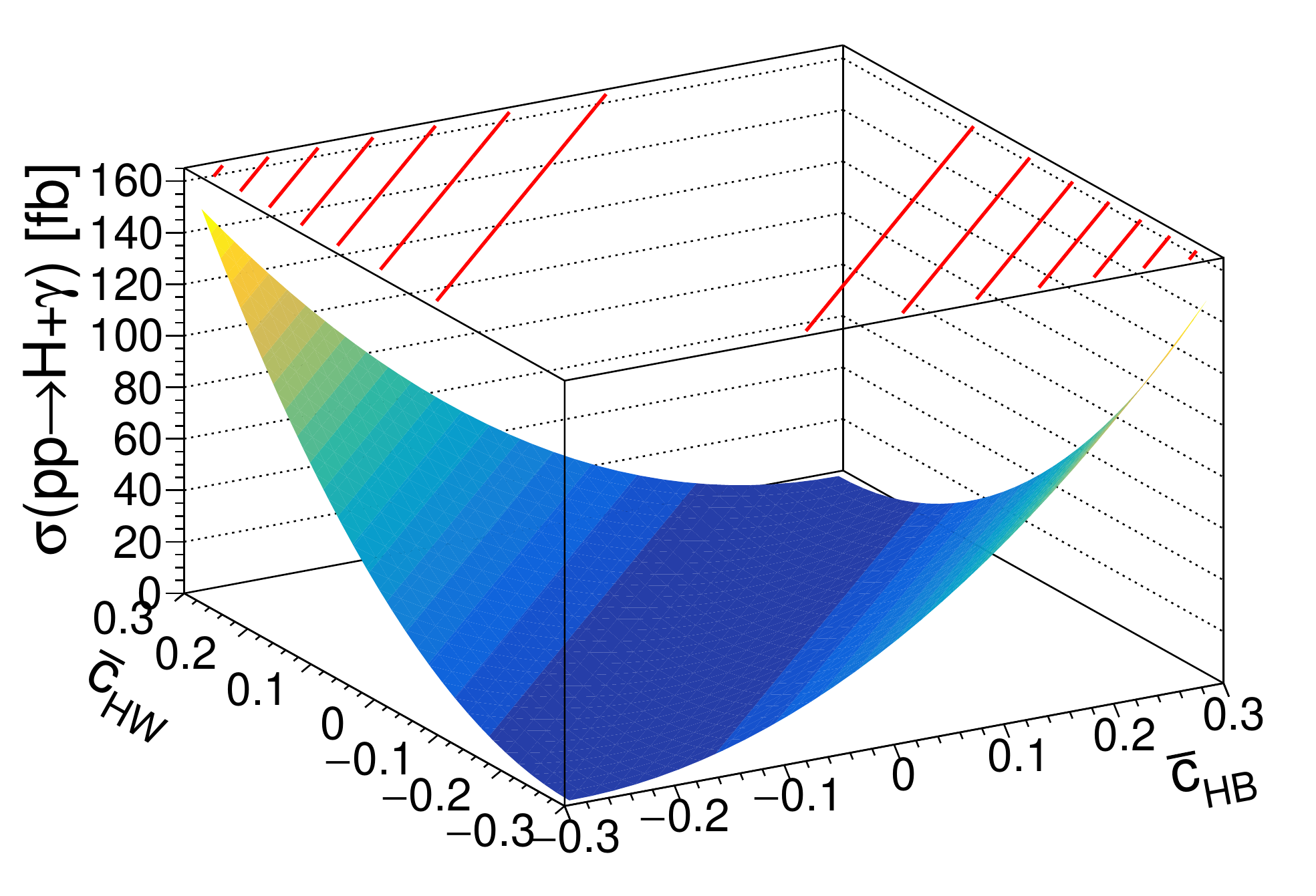}
\caption{\label{fig:xsec_2d} Two-dimensional parametrization of the signal cross section of the \Hg process with different values of \ca, \chw and \chb.
Apart from the two parameters indicated in each plot, the third parameter is fixed to 0.}
\end{figure}

%%%%%%%%%%
\section{Analysis strategy}
\label{sec:analysis}
%%%%%%%%%%

The ATLAS \Hg resonance search~\cite{Aaboud:2018fgi} is carried out to search for heavy resonances decaying to the SM Higgs boson and a photon, using the $b\bar b$ decay of the Higgs boson.
In its signal region, both the selected photon and the Higgs boson are highly boosted (with large momenta).
The search is performed by looking for a bump in the smooth background of the \Hg invariant mass spectrum $m_{H\gamma}$. As reported in the ATLAS paper, the mass spectrum observed is consistent with the background-only hypothesis and no evidence of new resonances is found.

The highly boosted signature is of particular interest for probing the anomalous Higgs-photon coupling, 
as the BSM signal contribution may show longer tails extending up to the TeV scale in the $\mHg$ and photon \pT distributions, while the SM expectation drops more steeply~\cite{Khanpour:2017inb}.
Instead of performing a bump hunt on the $m_{H\gamma}$ spectrum as in the original ATLAS paper, we perform a counting experiment with the published $m_{H\gamma}$ spectrum to constrain the anomalous coupling of the Higgs boson. According to the ATLAS paper~\cite{Aaboud:2018fgi}, 138 events were observed in the signal region $800~\GeV<m_{H\gamma}<3.2~\TeV$, consistent with the expected number of background events $138\pm12$. 
We reinterpret the ATLAS data as follows.

The expected number of events in the signal region can be expressed as $s+b$, where $s$ and $b$ are the expected number of signal and background events, respectively. To constrain the Wilson coefficients, we construct a likelihood function assuming that the number of observed events $n$ follows a Poisson distribution with an expectation value $s+b$:
\begin{eqnarray}
  {\cal L} = \textrm{Pois}(n|s+b) \times \textrm{Gaus}(b_0| b, \sigma_b)\ .
\end{eqnarray}
Here $b$ is treated as a nuisance parameter. It is constrained by a Gaussian term with a mean value $b_0$ and a standard deviation $\sigma_b$. Both $b_0$ and $\sigma_b$ are obtained from the background fits in the ATLAS \Hg paper~\cite{Aaboud:2018fgi}. The expected number of signal events $s$ depends on the Wilson coefficients $\bar c_i$. It can be further expressed as:
\begin{equation}
  s = L_\textrm{int} \times \varepsilon \times \sigma(\bar c_i) \times Br\ ,
\end{equation}
The integrated luminosity $L_\textrm{int}$ of the ATLAS data sample is 36.1~\ifb. The SM $H\rightarrow b\bar b$ branching ratio $Br=58\%$ for the 125 GeV Higgs boson~\cite{deFlorian:2016spz} is used.
The signal efficiency $\varepsilon$ accounts for the event loss due to detector effects, and to the reconstruction and selection efficiencies in the ATLAS analysis.
It is determined by applying the efficiency table published in the ATLAS \Hg paper~\cite{Aaboud:2018fgi} to the simulated \mHg spectra in the signal Monte Carlo samples, and allows to evaluate the overall efficiency.
The \Hg production cross section $\sigma(\bar c_i)$ is computed in terms of the Wilson coefficients \ca, \chw and \chb, as described in Section~{\ref{sec:eft}}.
The Wilson coefficients \ca, \chw and \chb are treated as the parameters of interest (POIs).

Constraints on the Wilson coefficients are obtained by evaluating the profiled likelihood ratio assuming the asymptotic approximation~\cite{Cowan:2010js}:
\begin{equation}
   \lambda(\bar c_i) = \frac {{\cal L}(\bar c_i,\hat{\hat b})} {{\cal L}(\hat{\bar{c_i}},\hat b)}.
\end{equation}
Here, the numerator is the conditional maximum-likelihood function, where $\hat{\hat b}$ is the value of the nuisance parameter $b$ that maximizes the likelihood function for a given set of values of the Wilson coefficients $\bar c_i$. 
The denominator is the unconditional maximum-likelihood function where $\hat{\bar{c_i}}$ and $\hat b$ are the maximum-likelihood estimates of $\bar c_i$ and $b$, respectively.

%%%%%%%%%%
\section{Results and discussion}
\label{sec:result}
%%%%%%%%%%

A one-dimensional likelihood scan is performed to obtain constraints on each of the three Wilson coefficients in the EFT framework with the other two fixed to 0. The constraints on \ca, \chw and \chb are shown in Figure~\ref{fig:likelihood_1d_ca}. The 68\% and 95\% confidence intervals are shown in Table~\ref{tab:stat_interval}.

\begin{figure}[b]
\includegraphics[width=0.32\textwidth]{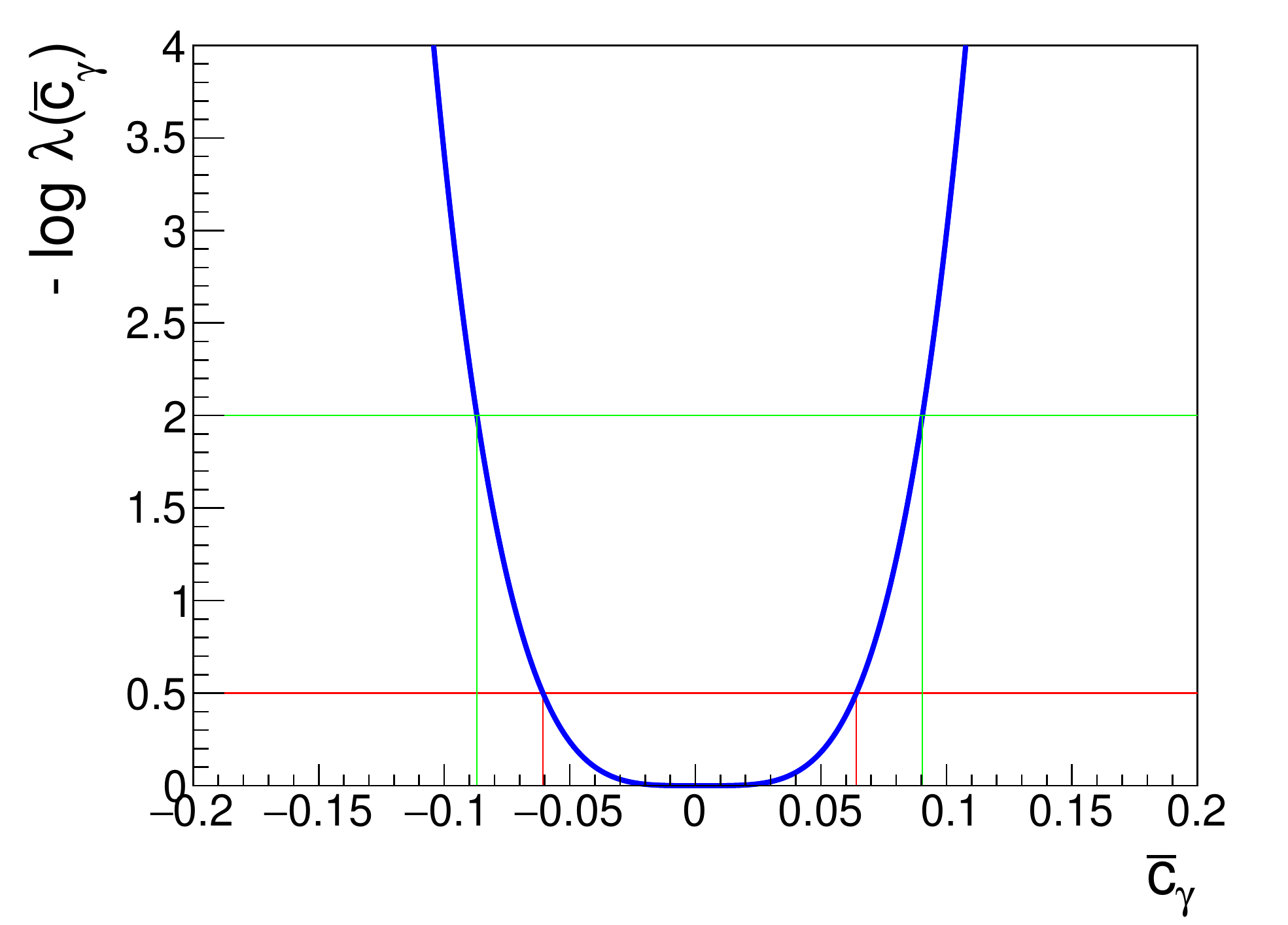}
\includegraphics[width=0.32\textwidth]{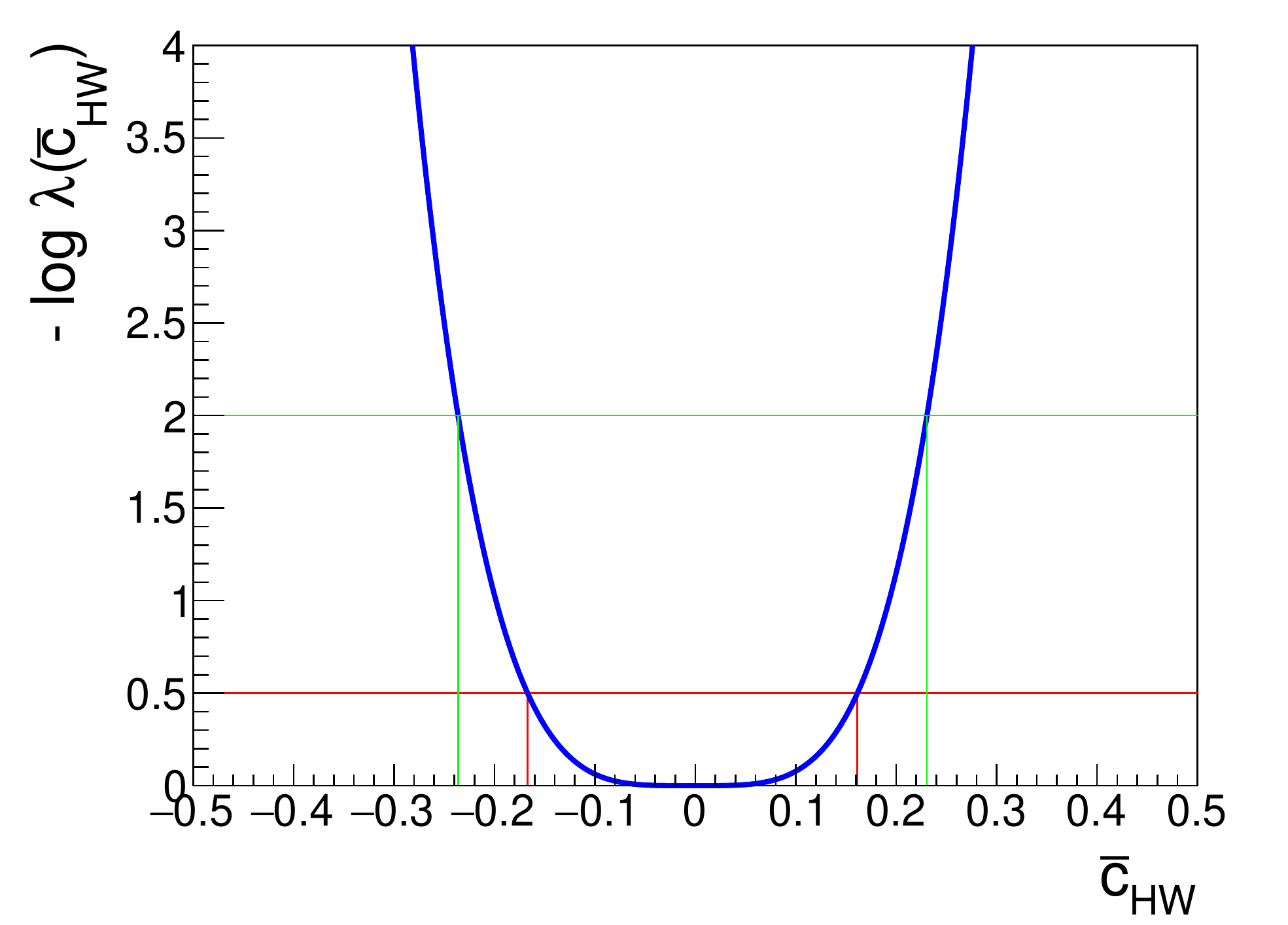}
\includegraphics[width=0.32\textwidth]{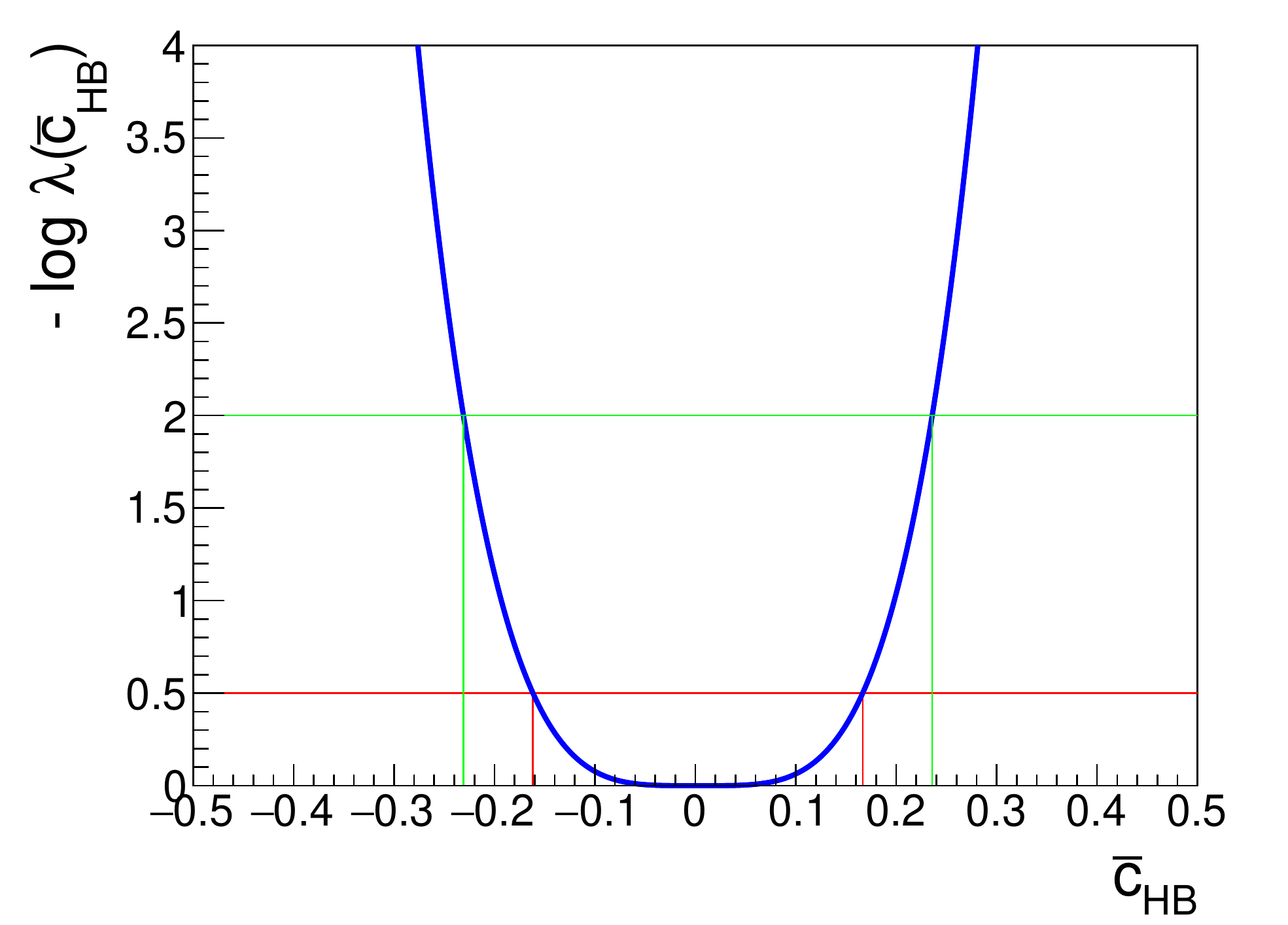}
\caption{\label{fig:likelihood_1d_ca} One-dimensional likelihood scan of the Wilson coefficients \ca (left), \chw (middle) and \chb (right) in the EFT framework with all the other coefficients fixed to 0. The 95\% (68\%) confidence interval is indicated by the red (green) line. The constraints are obtained from the data in the mass range 800~\GeV $< \mHg <$ 3.2~\TeV.}
\end{figure}

\begin{table}[hbt]
\caption{\label{tab:stat_interval}The 68\% and 95\% confidence intervals for the Wilson coefficients \ca, \chw and \chb in the EFT framework.}
%\begin{ruledtabular}
\begin{tabular}{ccc}
\hline
 Parameter & 68\% C.L. & 95\% C.L. \\
 \colrule
\ca   & [-0.061, 0.064] & [-0.087, 0.090] \\
\chw  & [-0.167, 0.161] & [-0.236, 0.231] \\
\chb  & [-0.162, 0.167] & [-0.230, 0.236] \\
\hline
\end{tabular}
%\end{ruledtabular}
\end{table}

Two-dimensional likelihood scans are also performed and the confidence regions are shown in Figure~\ref{fig:likelihood_2d}. Apart from the two Wilson coefficients indicated in the plot, the remaining one is fixed to 0 during the scan.

\begin{figure}[b]
\includegraphics[width=0.32\textwidth]{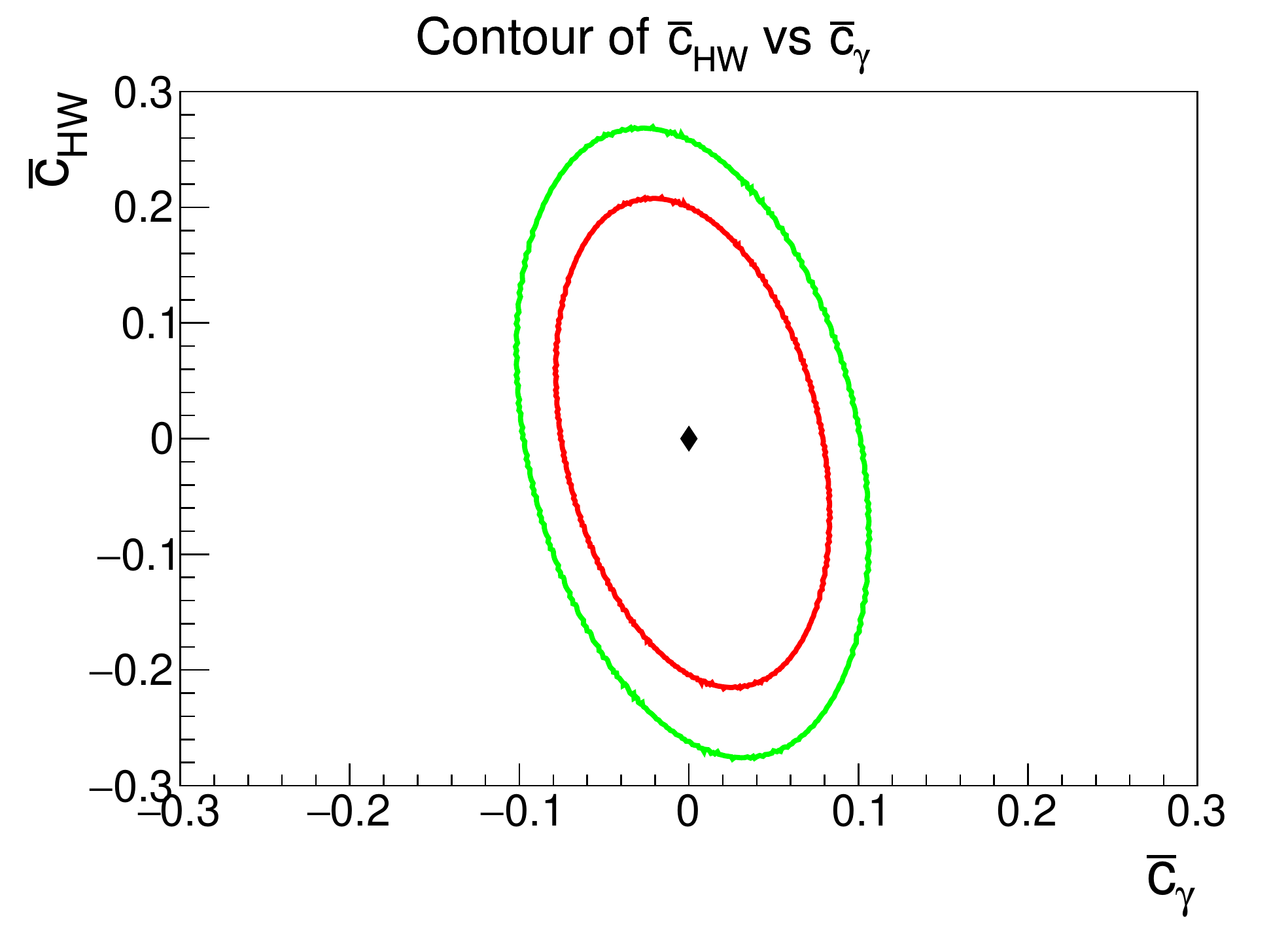}
\includegraphics[width=0.32\textwidth]{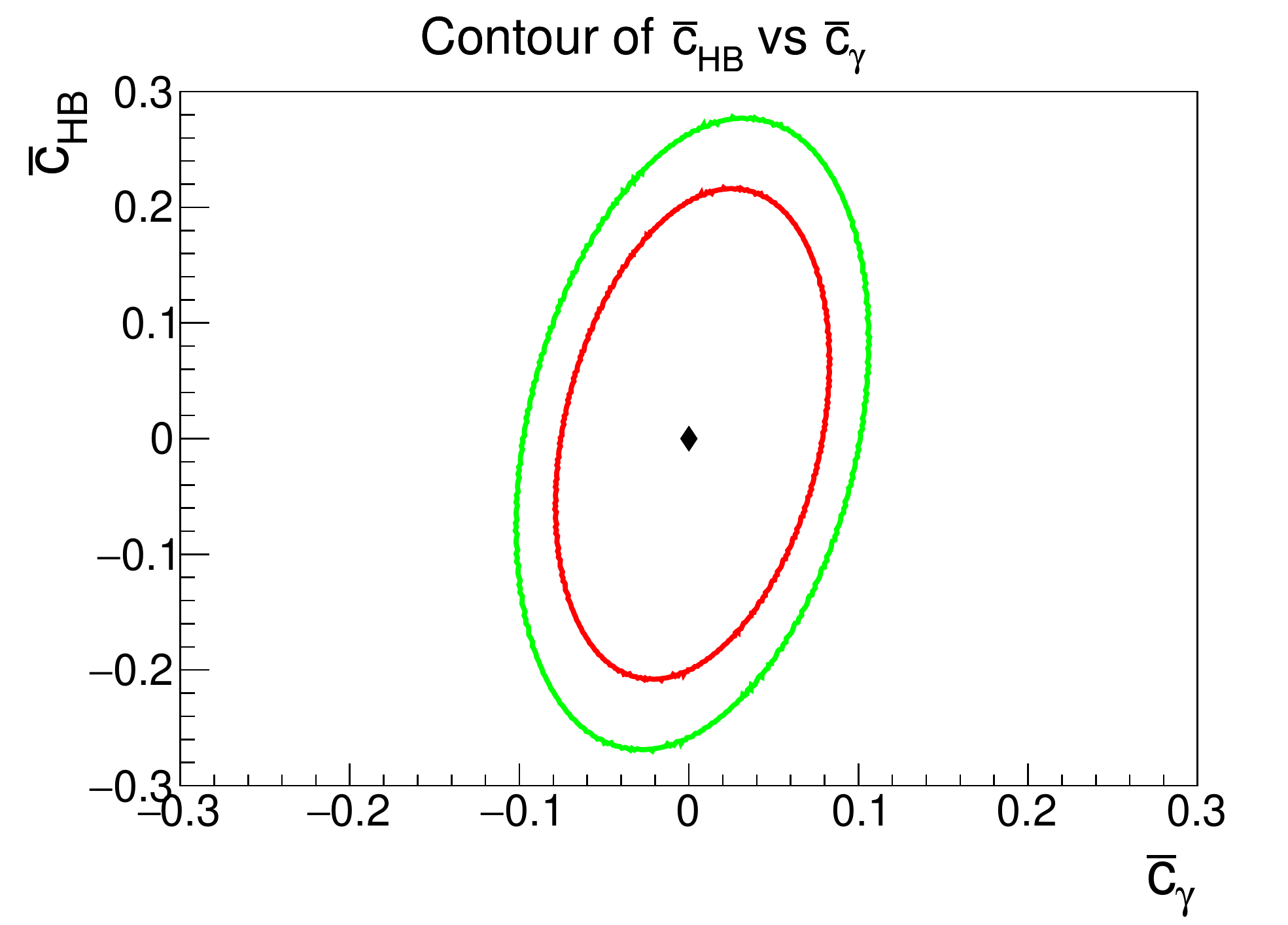}
\includegraphics[width=0.32\textwidth]{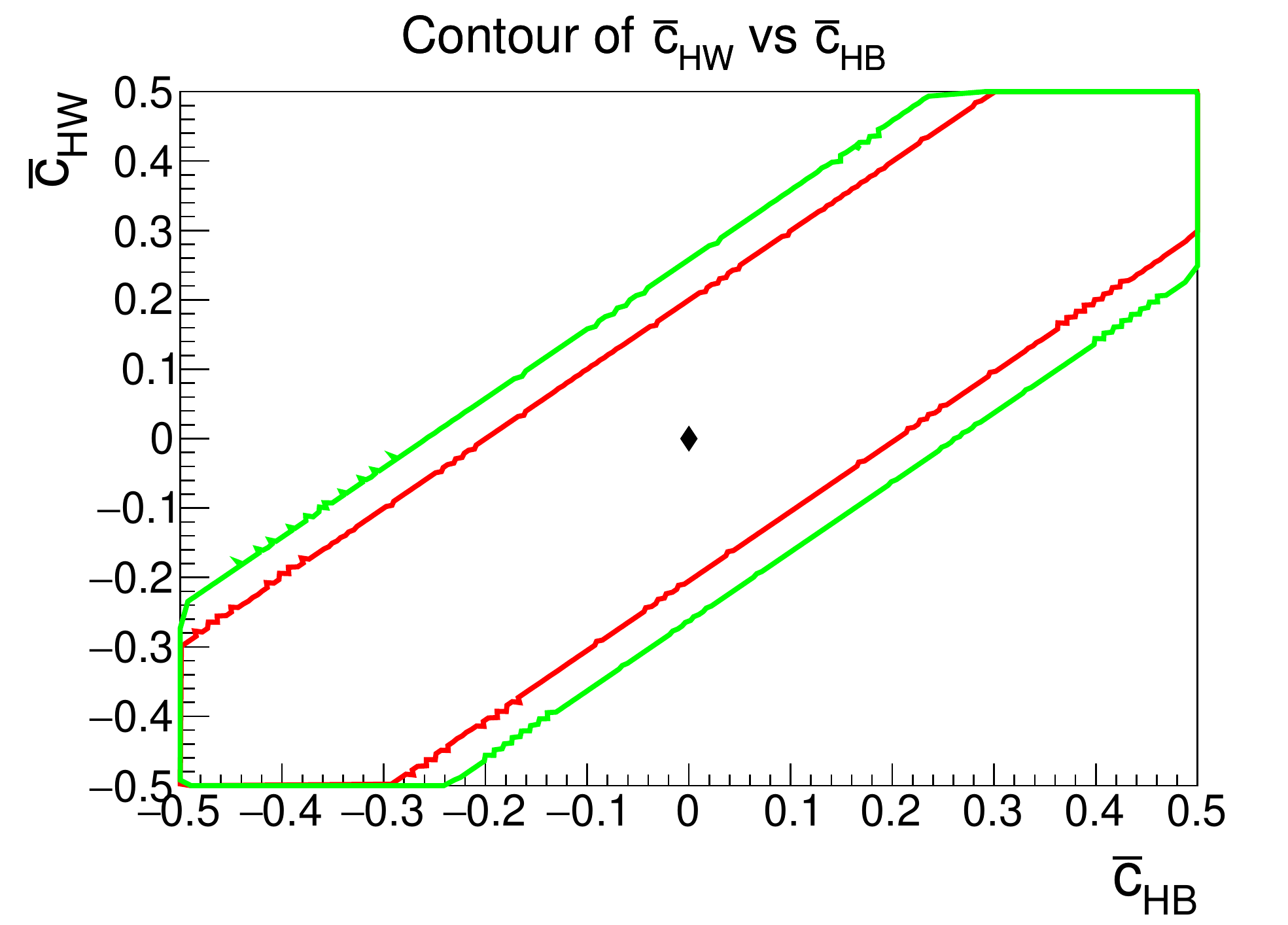}
\caption{\label{fig:likelihood_2d} Two-dimensional likelihood scan of the Wilson coefficients in the EFT framework. Apart from the two parameters indicated in each plot, the third parameter is fixed to 0. The 95\% (68\%) confidence region is indicated by the red (green) contour. The constraints are obtained from the data in the mass range 800~\GeV $< \mHg <$ 3.2~\TeV. The SM expectation at (0, 0) is also shown.}
\end{figure}

We compare the \Hg channel results with those obtained in the combined $H \to \gamma\gamma$ and $H \to ZZ^{*} \to 4\ell$ channels based on the same $pp$ dataset collected by the ATLAS experiment. The 68\% C.L. intervals from the combined channels~\cite{ATL-PHYS-PUB-2017-018} are:
\begin{eqnarray}
    \ca &\in& [-1.5\times 10^{-4},\ 2.2 \times 10^{-4}] ,\nonumber\\
    \chw &\in& [-0.080,\ -0.024] ,\nonumber\\
    \chb &\in& [-0.051,\ 0.103].\nonumber
\end{eqnarray}
While the limit on \ca is much stringent, the limits on \chw and \chb are of the same order of magnitude as the \Hg channel results. These results demonstrate excellent sensitivity of the \Hg production process to some of the Wilson coefficients in the EFT framework. The combination of the \Hg channel with the other channels could further improve the sensitivity for the Higgs boson anomalous couplings.

The limits can be further improved by considering the shape information from the differential distributions instead of doing a simple counting experiment.
In the $H\to\gamma\gamma$ channel, the 95\% C.L. observed limit for \chw has been improved to $-0.057<$\chw$<0.051$ after including differential distributions~\cite{Aaboud:2018xdt}, which is four times better than the limit achieved from the \Hg channel in our study.
In the \Hg channel, improvements to the sensitivity are also anticipated by including additional information from the $\mHg$ and photon \pT distributions, 
but we consider the shape analysis beyond the scope of this paper.

%%%%%%%%%%
\section{Conclusions}
\label{sec:conclusion}
%%%%%%%%%%
We present an interpretation of the recent ATLAS \Hg resonance search results with 36.1~\ifb of $pp$ collision data at $\sqrt{s}=13$ TeV in view of the search for the Higgs boson anomalous coupling in the \Hg final state. We provide constraints on the Wilson coefficients of dimension-six EFT operators for the first time in the \Hg final state with $pp$ collision data. These results demonstrate excellent physics potential of the \Hg production process. With differential cross sections measured for the \Hg process in the future, the constraints can be further improved.

\bibliographystyle{unsrt}
\bibliography{hginterpret}% Produces the bibliography via BibTeX.

\end{document}
%
% ****** End of file apssamp.tex ******